\documentclass[useAMS]{mn2e}

\usepackage{graphicx}
\usepackage{amssymb}

\font\tenbg=cmmib10 at 10pt

\def \rvecphi{{\hbox{\tenbg\char'036}}}

\font\tenbg=cmmib10 at 10pt

\def \rvecphi{{\hbox{\tenbg\char'036}}}

\def\lesssim{\mathrel{\hbox{\rlap{\hbox{\lower4pt\hbox{$\sim$}}}\hbox{$<$}}}}
\def\gtrsim{\mathrel{\hbox{\rlap{\hbox{\lower4pt\hbox{$\sim$}}}\hbox{$>$}}}}

\title{Transmission Line Analogy for Relativistic Poynting-Flux Jets}

\author[R.V.E. Lovelace \& P.P. Kronberg]{R.V.E. Lovelace$^{1}$ \& P.P. Kronberg$^{2, ~\!3}$\\
$^{1}${Department of Astronomy, Cornell University, Ithaca, NY
14853; email: lovelace@astro.cornell.edu} \\
$^{2}${Theoretical Division, Los Alamos Nat. Lab, Los Alamos, NM 87545 and} \\
$^3${Department of Physics, University of Toronto, ON M5S 1A7, Canada}\\
}

\begin{document}

\maketitle

\begin{abstract}

    Radio emission, polarization,
and Faraday rotation maps of the radio jet of the galaxy 3C 303 have
shown that one knot of this jet carries a {\it galactic}-scale electric  current and that it is magnetically dominated.
    We develop the theory of magnetically dominated or Poynting-flux jets by  making an  analogy of a Poynting jet with
 a transmission line or waveguide carrying a net current and
 having a  potential drop across it (from the jet's  axis to its radius)
  and a definite impedance  which we derive.
    The electromagnetic energy flow in the
 jet is the jet impedance times the square of the jet current.
   The observed current in 3C 303 can be used to 
calculate the electromagnetic
energy flow in this magnetically dominated jet.  
    Time-dependent but not necessarily small
perturbations of a Poynting-flux jet 
are described by the  ``telegrapher's equations.''   These
predict the propagation speed of disturbances and the 
effective wave impedance for forward and backward
propagating wave components.  
   A  localized disturbance of a  Poynting jet  gives rise
to localized dissipation in the jet which may explain the 
enhanced synchrotron radiation in the knots of the
3C 303 jet, and also in the apparently stationary knot HST-1
in the jet near the nucleus of the nearby galaxy M87. 
   For a  relativistic Poynting jet on parsec scales,
the reflected voltage
wave from an inductive termination or load can lead to a backward
propagating wave which breaks down the magnetic insulation
of the jet giving $|{\bf E}| /|{\bf B}|\geq 1$.   
   At the threshold for breakdown, $|{\bf E}|/|{\bf B}|=1$, 
positive and negative particles are directly accelerated 
in the ${\bf E \times B}$ direction which is approximately
along the jet axis.   Acceleration can occur up 
to Lorentz factors $\sim 10^7$.     This particle acceleration mechanism is distinct from that in  shock waves and that in magnetic field
reconnection.

 \end{abstract}

\section{Introduction}

	A fundamental open question of
astrophysical jet models is how energy 
extracted from the accretion flow close to the black hole event horizon
is transported along narrow jets to much larger distances.
   The total energy carried by the jets of active galaxies  is estimated to be 
a non-negligible fraction of the massive black hole formation energy, $\sim 0.1 M_{\rm bh}c^2$ (Kronberg et al. 2001).   
       The jets  are initially highly relativistic and
low-density and for this reason are {\it magnetically dominated} or {\it force-free}.  In this limit
a negligible fraction of the power is carried by the particle kinetic
energy.
   That is,
the energy outflow from the accretion disc is
 in the form of a collimated
 `Poynting-flux jet'  as proposed by Lovelace (1976) 
and subsequently studied in many papers  
(Benford 1978; Lovelace, Wang, \& Sulkanen 1987; Lynden-Bell 1996; 
Li, et al. 2001; Lovelace, et al. 2002; Lovelace \& Romanova 2003; Nakamura et al. 2008).

    An illustrative model for the formation of a relativistic
Poynting jet is sketched in Fig. 1.   
    A  large scale magnetic field is assumed to thread the accretion disc.
As a result of the differential rotation of the disc
the initial field loops opens up giving a current outflow (or inflow) $I$ along the spine 
 of the jet of cylindrical radius $r_J$ initially of the order of
 the Schwarzschild radius of the black hole (Lovelace \& Romanova
 2003).    A three-dimensional view of a 
 numerically calculated magnetic field is shown
 in Fig. 6 of Lovelace et al. (2002).
 In this model the jet power comes  from the accretion disc.
    The associated toroidal magnetic field $B_\phi$ is responsible for collimating the jet.
   An equal but opposite ``return current'' flows 
inward (or outward) at much larger radial distances 
from the jet axis so that the net current  outflow from the source is zero. 
      Because the jet current and  the return current have opposite signs 
they repel as a result of their magnetic interaction mediated by the toroidal magnetic field. 
      This repulsion between the jet and
its  return current  has been demonstrated in  magnetohydrodynamic
(MHD) simulations 
(Ustyugova et al. 2000; Nakamura et al. 2008).   
    The energy flow in the jet is unidirectional and is carried predominantly by the Poynting flux $\hat{\bf z}\cdot({\bf E\times B})/\mu_0 = E_r \times B_\phi /\mu_0$ (MKS units), where $E_r$ is the radial electric field of the jet.  
        The  power in the Poynting jet may come predominantly from the spin-down of the black  hole (Blandford \& Znajek 1977).
  This process, which depends on the presence of a magnetized  accretion
disc around the rotating black hole, has been studied with general relativisitic MHD simulations  (e.g., McKinney 2006; Beckwith, Hawley,
\& Krolik 2008;   Tcchekhovskoy, Narayan, \& McKinney 2011).

        Rotation measure gradients observed on parsec 
 scales close to the nuclear central black hole have 
given evidence of electric current flow along the
jets  (Asada {\it et al.} 2002; Gabuzda, Murray, 
\& Cronin 2004; Zavala \& Taylor 2005).   
     More recently,
on a scale $10^4$ times larger,
 radio emission, polarization, X-ray,
and Faraday rotation maps  
of the radio jet of the galaxy 3C 303 have
shown that one knot of this jet has a {\it galactic}-scale 
current  of $\sim 3\times 10^{18}$ Amp\`ere flowing along the jet axis (Kronberg et al. 2011).   
     The physical parameters of this knot derived from
the observations indicate that the jet is
magnetically dominated, that is, a Poynting jet.
   The existence of an axial current in this very large scale
plasma flow suggests an electrical circuit analogy for the
jet. 

    A further possible indication of a Poynting jet is
the jet near the nucleus of the galaxy M87 which has  been
observed to have an apparently stationary but strongly
time-dependent knot HST-1 in its jet
 (Biretta et al. 1999).    This knot showed
 a dramatic peak in emission in about 2005.2 in the
 radio, optical, and X-ray bands (Harris et al. 2009).

       In Sec. 2 we model a Poynting jet as a  transmission line carrying
a net axial current and  having   a potential drop  across it.
   Further, we derive the transmission line impedance and
the electromagnetic energy flow in the jet.
    Time-dependent but not necessarily small
perturbations of a Poynting-flux jet (\S 2.2) 
are described by the  ``telegraphers' equations''  (Heaviside 1893).  
These are wave equations for the current and voltage across the line.            The voltage and current consist in general of forward and backward propagating components.
    A localized irregularity of a  Poynting jet  (\S 2.3) can
give rise to localized dissipation in the jet which may explain the 
enhanced synchrotron radiation in the knots of the
3C 303 jet and in the apparently stationary knot HST-1 in M87.

  Secs. 2.4 \& 2.5 consider  highly relativistic Poynting jets on 
parsec-scales jets such as HST-1.   In this case the reflected voltage
wave from an inductive load can lead to a backward
propagating wave which breaks down the magnetic insulation
of the jet,  giving $|{\bf E}| / |{\bf B}|\geq 1$.   
      The threshold for this breakdown is $|{\bf E}|/|{\bf B}|=1$.
  At this threshold,
positive and negative particles are directly accelerated 
in the ${\bf E \times B}$ direction which is approximately
along the jet axis and can be accelerated to very high
 Lorentz factors.   
Sec. 3 gives conclusions of this work.

\begin{figure}
\centering
\includegraphics[scale=0.4]{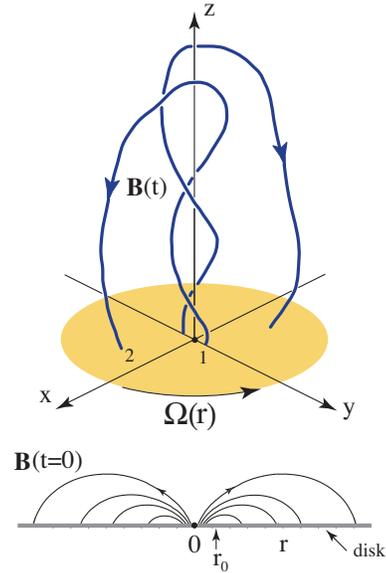}
\caption{Sketch of the magnetic
field configuration of a Poynting jet
adapted from Lovelace and Romanova (2003).
  The bottom part of the figure shows
the initial  dipole-like 
magnetic field threading the disc
which rotates at the angular rate
$\Omega(r)$. }
\end{figure}
     
\section{Theory}

    In cylindrical $(r,\phi,z)$
coordinates with
axisymmetry assumed,
    the magnetic field has
the form $ {\bf B}~ = {\bf B}_p +
B_\phi \hat{\rvecphi~},$ with $
{\bf B}_p = B_r{\hat{\bf r}}+
B_z \hat{\bf z}$, and
$B_r =-(1 / r)(\partial \Psi/ \partial z),$ 
and $B_z =(1 / r)(\partial \Psi / \partial r).$
Here,  $\Psi(r,z) \equiv r A_\phi(r,z)$ 
is the flux function.
  A simple form of this function is 
$\Psi(r,0)=(1/2){r^2B_0 /[1+2(r/r_0)^3]}, $
where $B_0$ is the axial magnetic field strength
in the center of the disc, and $r_0$ is
the radius of the $O-$point of the magnetic
field in the plane of the disc as indicated
in Fig. 1.    
    This $\Psi$ is taken to apply for
$r \geq 0$ even though it is not valid
near the horizon of the black hole. 
    The contribution
from this region is negligble for the
considered conditions where $(r_g/r_0)^2 \ll 1$,
with $r_g\equiv GM/c^2$.
     For a corotating disc around a Kerr black
hole the disc's angular velocity viewed from a large distance is
$\Omega ={[c^3/(GM) ]/[ a_*+(r/r_g)^{3/2}]},$
for $r>r_{\rm ms}$ where $r_{\rm ms}$ is the innermost
stable circular orbit and
$a_*$ is the spin parameter of the black hole
with $0 \leq a_* <1$. 

   At large distances from the disc ($z \gg r_0$) 
the flux function solution of the `force-free' 
Grad-Shafranov equation is 
\begin{equation}
\bar \Psi =  {\bar r^{4/3} \over 
[2 {\cal R}~(\Gamma^2 -1)~]^{2/3}} ~, 
\end{equation}
(Lovelace \& Romanova 2003).  
     The `force-free' solution is applicable
for conditions where the kinetic energy-density
of the plasma is much less than the magnetic
energy-density. 
     Here $\Gamma$ is
the Lorentz factor of the jet,
 $\bar r \equiv r/r_0$,
$\bar \Psi \equiv  \Psi/\Psi_0$ with $\Psi_0 \equiv r_0^2 B_0/2$,
and ${\cal R} \equiv r_0/r_g$.
  This dependence holds for
$\bar r_1 \equiv {[2(\Gamma^2-1)]^{1/2}/{\cal R}} < 
\bar r < \bar r_2 =
{[2 {\cal R}(\Gamma^2-1)]^{1/2}/ 3^{3/4}}.$
At the inner radius $\bar r_1$, $\bar \Psi = 1/{\cal R}^2$,
which corresponds to the streamline which passes
through the disc at a distance $r=r_g$.
   For $\bar r < \bar r_1$, we assume
$\bar \Psi \propto \bar r^2$, which
corresponds to $B_z=$ const.
   At the outer radius $\bar r_2$,
$\bar \Psi =(\bar \Psi)_{\rm max} =1/3$ which
corresponds to the streamline which goes
through the disc near the $O-$point at
$r=r_0$.   
    Note that there is an appreciable range of radii
if ${\cal R}^{3/2} \gg 1$.
   Also note that for $z \gg r_0$,  radiation drag
on the lepton component of the jet is negligible.

  For $\bar{r}_1 <r < \bar{r}_2$,  the
field components of the Poynting jet are
$$
\bar E_r = -\sqrt{2}~(\Gamma^2-1)^{1/2}~ \bar B_z~,~~~
\bar B_\phi = -\sqrt{2}~\Gamma~\bar B_z~,
$$
and
\begin{equation}
\quad\quad ~\bar B_z = {2\over 3} {\bar r^{-2/3} 
\over[2{\cal R}(\Gamma^2-1)]^{2/3}}~.
\end{equation}
This electromagnetic field satisfies
the radial force balance equation,
$ {d B_z^2 / dr}+ (1/ r^2){d [ r^2(B_\phi^2 -E_r^2)] /dr} =0$,
for an axisymmetric, translationally-symmetric, and time-independent
force-free field.  Figure 2 shows the field profiles.

   Appendix A describes a different force-free jet model with
electric and magnetic fields similar to those in a common coaxial
transmission line.

   At the jet radius $r_2$,
there is a boundary layer where
the axial magnetic field changes from $B_z(r_2-\varepsilon)$
to zero at $r_2+\varepsilon$,  where $\varepsilon \ll r_2$ is the
half-width of this layer.
     The electric field
changes from $E_r(r_2-\varepsilon)$ to zero at $r_2+\varepsilon$.
    The  toroidal magnetic field
changes from $B_\phi(r_2-\varepsilon)$ to $B_\phi(r_2+\varepsilon)$
where this change is fixed by
the radial force balance.
   Thus for $r>r_2$, we have 
$E_r = 0$, $B_z=0$, and 
$ B_\phi = \sqrt{3} B_z(r_2-\varepsilon)(r_2/r)=
\sqrt{(3/2)}\Gamma^{-1}B_\phi(r_2-\varepsilon)(r_2/r)$.
Equivalently, $B_\phi(r_2+\varepsilon)/B_\phi(r_2-\varepsilon)
=\sqrt{3/2}/\Gamma$.

      The toroidal magnetic field for $r>r_2$ applies out to an
`outer radius' $r_3$ where the magnetic pressure of the jet's
toroidal magnetic field, $B_\phi^2(r_3)/8\pi = p_{\rm ex}$,
balances the external ram pressure $P_{\rm ex} = p_{\rm ex}
+\rho_{\rm ex} (dr_3/dt)^2$, where 
$p_{\rm ex}=n_{\rm ex}k_{\rm B}T_{\rm ex}$ is the kinetic
pressure of the external intergalactic plasma and $\rho_{\rm ex}$
is its density.   The outward propagation of the jet will be
accompanied by the non-relativistic expansion of the outer radius,
$dr_3/dt >0$.

      We take as the `jet current' the axial current $I_0$
flowing along the jet core $r\leq r_2-\varepsilon$.
  From  Amp\`ere's law, $B_\phi(r_2)  =-2 I_0/(c~\!r_2)$ 
  or in convenient units, $B_\phi[{\rm G}] = -I_0[{\rm A}]/(5 r[{\rm cm}])$.  
  The net current carried by the jet ($r\leq r_2+\varepsilon$)
  is $I_{\rm net} = \sqrt{3/2}\Gamma^{-1} I_0$.
  
    Using equations (2), the energy flux carried by the Poynting jet 
can be expressed as
 \begin{equation}
 \dot{E}_J = \left[{c\over 2},~{2\pi \over \mu_0}\right]
 \int_0^{r_2} r dr E_r B_\phi = {\cal Z}_0I_0^2~,
 \end{equation}
 where the square bracket is for cgs or MKS units,  and
 \begin{equation}
 {\cal Z}_0= {3\over c}\beta
 ~[{\rm cgs}]
 =90\left(1-{1\over \Gamma^2}\right)^{1/2}~\Omega~[{\rm MKS}]~,
\end{equation}
where $\beta=U_z/c = (1-\Gamma^{-2})^{1/2}$.
Here, ${\cal Z}_0$ is the DC impedance of the Poynting jet.
 The conversion to MKS units is 
$c^{-1} \rightarrow (4\pi)^{-1}(\mu_0/\epsilon_0)^{1/2} =  30~\! \Omega.$
Earlier, the impedance of a relativistic Poynting jet was estimated
to be $\sim c^{-1}$ (Lovelace 1976).   

    We can apply these concepts to the jet in 3C 303 where
the observed axial current in the E3 knot is $3.3\times 10^{18}$ A 
(Kronberg et al. 2011),    Thus the electromagnetic energy 
flux is $\dot{E}_J \approx 8\times 10^{45}\beta$ erg s$^{-1}$.      
   This energy flux is much larger than the photon 
luminosity of the jet of $3.7\times 10^{41}$ erg s$^{-1}$
integrated over $10^8$ to $10^{17}$ Hz (Kronberg et al. 2011) 
assuming $\beta=U_z/c$ is not much smaller than unity.   
    For the E3 knot the jet radius is $r_2 \approx 0.5$ kpc 
so that $B_\phi(r_2) \approx 0.4$ mG.     
   The E3 knot is about $17.7$ kpc in projection from the galaxy nucleus.

\subsection{Transmission Line Analogy}

     Here, we interpret the Poynting flux jet   in terms of a transmission
line  as proposed earlier  by Lovelace \& Ruchti (1983).
  The different physical quantities are measured in
the `laboratory' frame which is the rest  frame of the plasma
outside of the jet at $r \geq r_2$.
      The effective potential drop across 
the transmission line  is taken to be 
\begin{equation}
V_0 = - {1\over 2} r_0 \int_0^{\bar{r}_2} d\bar{r} E_r(\bar r) ~~=~~ {r_0 \over 3^{1/4}}
{B_0 \over \sqrt{{\cal R}}}~,
\end{equation}
where the factor of one-half accounts for the fact
that the transmission line does not consist of two
conduction surfaces.

    The axial current flow of the jet is 
\begin{equation}
I_0= - {1\over 2} c r_2 B_\phi(r_2) ={ V_0 \over {\cal Z}_0}~,
\end{equation}
with ${\cal Z}_0$ given by equation (4).
The units of equations (4) and (5) are cgs.
    In MKS units note that a current $I_0 = 3\times 10^{18}$ A
gives a voltage $V_0 = 2.7\times 10^{20}\beta$ V.
    The energy of ions accelerated across this voltage is 
large enough  to account for ultra high energy cosmic
rays (Lovelace 1976;  Biermann, Kang, \& Ryu 2001; Ostrowski 2002).

\begin{figure}
\centering
\includegraphics[scale=0.45]{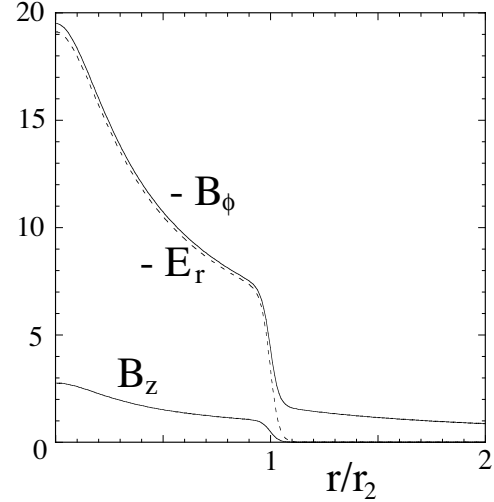}
\caption{Radial dependence of the field components of
a Poynting jet from equation (8).  The inner jet radius is
assumed to be $r_1 =0.05 r_2$, the boundary layer
thickness is $\varepsilon=0.01r_2$, and $\Gamma=5$.}
\end{figure}

 There is a corresponding electric field energy per unit length of the
jet which in MKS units is
\begin{equation}
w_E ={\epsilon_0 \over 2}2\pi \int_0^{r_2} r dr E_r^2~ = ~{1\over 2}C V_0^2~.
~~~~{\rm Here,}~
C={4\pi\epsilon_0 \over 3}~
\end{equation}
is the capacitance per unit length in Farads per meter and
$\epsilon_0 =8.854\times 10^{-12}$ F/m.

    The magnetic energy per unit length of the jet
in MKS units is
\begin{eqnarray}
w_B&=&{\pi\over  \mu_0}\int_0^{r_2} r dr (B_\phi^2 + B_z^2)+
{\pi \over \mu_0} \int_{r_2}^{r_3} r dr B_{\phi+}^2 \left({r_2 \over r}\right)^2~,
\nonumber \\
&=&{1\over2}L I_0^2~.
\end{eqnarray}
Here, $B_{\phi+}$ is the toroidal field at $r_2+\varepsilon$.
     Carrying out the integrals we find
\begin{equation}
L ={3 \mu_0 \over 4 \pi}\left[1+{1\over 2 \Gamma^2}
+{1\over 2\Gamma^2}\ln\left({r_3 \over r_2}\right)\right]~,
\end{equation}   
which is the inductance per unit length in Henrys per meter
with $\mu_0=4\pi \times 10^{-7}$ H/m.

\subsection{Telegraphers' Equations} 

Time and space ($z-$)dependent  
perturbations (not necessarily small) of a Poynting-flux jet are described by the Telgrapher's equations,
\begin{equation}
{\partial \Delta V \over \partial t}= - {1\over C} {\partial \Delta I \over \partial z}~,
\quad \quad{\partial \Delta I \over \partial t}= - {1\over L} {\partial \Delta V \over \partial z}~,
\end{equation}
where $(\Delta  V, ~\Delta I)$ represent deviations from the equilibrium
values $(V_0,~I_0)$ (Heaviside 1893; Bergeron 1977; Samokhin 2010).
The equations can be combined to give the wave equations
\begin{equation}
\left({\partial^2 \over \partial t^2} -u_\varphi^2{\partial^2\over \partial z^2}\right)( \Delta V,~\Delta I) =0~,
\end{equation}
where
\begin{equation}
u_\varphi ={1\over \sqrt{LC}}~=~c\left[1+{1\over 2 \Gamma^2}
+{1\over 2\Gamma^2}\ln\left({r_3 \over r_2}\right)\right]^{-1/2}~,
\end{equation}
is the phase velocity of the perturbation.
   The general solution of equation (10) is
\begin{eqnarray}
\Delta V&=&\Delta V_+(z-u_\varphi t)+\Delta V_-(z+u_\varphi t)~,\nonumber \\
\Delta I&=&\Delta I_+(z-u_\varphi t)+\Delta I_-(z+u_\varphi t)~,
\end{eqnarray}
with
$$
\Delta V_+ = {\cal Z} \Delta I_+ \quad {\rm and} \quad  
\Delta V_-   = -{\cal Z} \Delta I_- ~.
$$
Here, 
\begin{equation}
{\cal Z} = \sqrt{L\over C} =90 \left[1+{1\over 2 \Gamma^2}
+{1\over 2\Gamma^2}\ln\left({r_3 \over r_2}\right)\right]^{1/2}\Omega ~[\rm MKS],
\end{equation}
is the wave impedance of the jet.

\subsection{Irregularities in the Transmission Line:}

    A localized irregularity may appear in the transmission line
due possibly to an instability at $t > 0$ as diagrammed in Fig. 3.
 The irregularity  can be modeled as an
 extra impedance ${\cal Z}_\ell$ or `load' across the transmission line at $z=0$.
     This impedance is considered
 to go from ${\cal Z}_\ell(t<0)=\infty$ to a constant value ${\cal Z}_\ell$ for $t>0$.
    In general ${\cal Z}_\ell$ is complex with an
imaginary component (reactance) which can be positive
(capacitive) or negative (inductive).
    On either side of the discontinuity the line is
assumed to have the impedance ${\cal Z}$  given by equation (14).
    On the upstream side of ${\cal Z}_\ell$ ($z<0$), the line voltage
is $V_0 + \Delta V_-$,  where $\Delta V_-$
is the  backward  propagating wave.   
     The current on this part of
the line is $I_0+\Delta I_-$. 
    There is no forward propagating wave for the 
conditions considered here.
   On the downstream side of ${\cal Z}_\ell$, the line voltage
is $V_0 + \Delta V_t$ and the current is $I_0+\Delta I_t$, where
$(\Delta V_t,~\Delta I_t=\Delta V_t/{\cal Z})$ represent the transmitted
wave.

\begin{figure}
\centering
\includegraphics[scale=0.6]{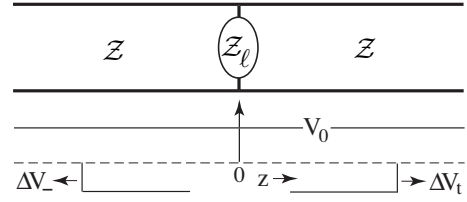}
\caption{Diagram of an irregularity in
the transmission line represented by
the impedance ${\cal Z}_\ell(t)$ where
${\cal Z}_\ell(t<0)=\infty$ and ${\cal Z}_\ell(t>0)$ is finite.  Here,
$V_-$ the wave reflected off of ${\cal Z}_\ell$, 
$V_t$ the transmitted wave, and $V_0$ is the line voltage for $t<0$.  }
\end{figure}

     The standard conditions on the potential 
and current flow at ${\cal Z}_\ell$ ($z=0$)
give $V_0+\Delta V_- = V_0+\Delta V_t$ and 
$I_0 +\Delta I_- = I_\ell +I_0 +\Delta I_t$, where $I_\ell=
(V_0+\Delta V_-)/{\cal Z}_\ell$
is the current flow through  ${\cal Z}_\ell$.
      In this way we find
 \begin{equation}
 \Delta V_- = {-{\cal Z} \over 2{\cal Z}_\ell +{\cal Z}}V_0~=~\Delta V_t~.
\end{equation}
Note that for ${\cal Z}_\ell \rightarrow \infty$,  both $\Delta V_- $
and  $\Delta V_t $ tend to zero.

     The power loss rate in the load ${\cal Z}_\ell$ is
\begin{equation}
\dot{\cal E}_\ell = {(V_0+\Delta V_t)^2 \over {\cal Z}_\ell}~=~ 
{4 {\cal Z}_\ell \over (2 {\cal Z}_\ell +{\cal Z})^2} V_0^2~.
\end{equation}
We assume that this power goes into accelerating charged particles
which in turn produce the observed synchrotron radiation.
   This power could account for the emission of the E3 knot of
3C 303 (Kronberg et al. 2011) and, on parsec scales, the emission of the HST-1 knot in the M87 jet (Biretta et al. 1999).

    The power $\dot{\cal E}_\ell$ is equal to the difference 
between the Poynting flux for $z<0$, 
\begin{equation}
{\cal S}^{-} = (V_0+\Delta V_-)(I_0+\Delta I_-)=
{[V_0^2 -(\Delta V_-)^2]\over{\cal Z}}~,
\end{equation}
and the Poynting flux for $z>0$, 
\begin{equation}
{\cal S}^+ =
(V_0+\Delta V_t)(I_0+\Delta I_t)={[V_0^2+2V_0\Delta V_t +(\Delta V_t)^2]\over {\cal Z}}.
\end{equation}
Using equation (17) we find that ${\cal S}^- -{\cal S}^+ = \dot{\cal E}_\ell$.

\begin{figure}
\centering
\includegraphics[scale=0.55]{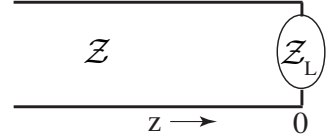}
\caption{Diagram of a transmission line of
impedance ${\cal Z}$ terminated by a  load
 impedance ${\cal Z}_L$ with an inductive component. 
 }
\end{figure}

\subsection{Magnetic Insulation}

      The Poynting jet is analogous  to
a magnetically insulated transmission line.
      In laboratory experiments magnetically 
insulated transmission lines involving
crossed electric and magnetic fields are commonly
used to transport
pulses of high energy content onto small targets (e.g., Shope et al. 1978; Samokhin 2010).
  The condition for ``magnetic insulation'' of a planar gap of width $d$
between a conducting cathode and anode with potential drop $V_0$
and initially filled with a uniform magnetic field $B_0$ is
$B_0^2 > (2V_eV_0+ V_0^2)d^{-2}$, where $V_e\equiv m_ec^2/e$ 
$\approx 5.14\times 10^5$ V
(Lovelace \& Ott 1974; Ron, Mondelli,
\& Rostoker 1973).

    In the astrophysical limit where $V_0 \gg V_e$,
the condition for magnetic insulation 
is $|{\bf E}| <|{\bf B}|$  for  the
case where field components have the same radial
dependence. 
    If the power flow in the jet  is {\it unidirectional},  as
assumed in equations (2), then the condition for magnetic
insulation is
\begin{equation}
{|{\bf E}| \over |{\bf B}|}
=\left({\Gamma^2-1 \over \Gamma^2+1/2}\right)^{1/2}  <1~~~{\rm or}~~~
{V_0 \over {\cal Z}_R I_0} <\left(1+{1\over2\Gamma^2}\right)^{1/2},
\end{equation}
where ${\cal Z}_R \equiv 90 \Omega$ is a reference impedance.
This inequality  is always satisfied because the fields are
time-independent and force-free ($\rho_e {\bf E} +{\bf J}\times {\bf B}/c=0$).

    Consider now the case where the power
flow is { \it not unidirectional} and where the fields are time
and space-dependent, i.e.,   $V(z,t)=\Delta V_+ +\Delta V_-$,~
$I(z,t)=\Delta I_+ +\Delta I_-$.
We assume $\Gamma^2 \gg1$ so that $(B_z/B_\phi)^2 = 
{\cal O}(\Gamma^{-2}) \ll 1.$  
      To a  first approximation the
electric and magnetic fields correspond to those in a 
common transverse electric and magnetic (TEM) mode in
a coaxial cable.  
   Unlike a coaxial cable where the two conducting surfaces
are separated by a dielectric, the Poynting jet  has distributed
charge and current densities. 
    Because $|E_r/B_\phi| = V/({\cal Z} I)$, 
the insulation condition $|{\bf E}| < |{\bf B}|$ can then
be written as
\begin{equation}
{V \over {\cal Z}  I}~ <~ {\cal C}~,
\end{equation} 
where
${\cal C} =1+{\cal O}(\Gamma^{-2})$ is a critical dimensionless
number slightly larger than unity.

\subsection{Consequences of Breakdown of  Magnetic Insulation}

     Here we discuss conditions where the magnetic insulation
breaks down in a localized region near the termination
of a Poynting jet due to a load ${\cal Z}_L$ which has
an inductive component (Fig. 4.).         
    An essential condition for this to occur is
that the wave propagation {\it not be unidirectional.}  
   The load is assumed to be stationary or 
sub-relativistic in the laboratory frame owing to 
the ram pressure from the external plasma.
   The breakdown of magnetic insulation
in laboratory transmission lines is discussed by
Gordeev (1978).

   The consequences of the breakdown can be dramatic: 
Nuclei can be accelerated to energies of the order
of  $e V \sim 3 \times 10^{20}$ eV while leptons
may be accelerated to smaller energies owing  to
radiative losses and interactions with radiation.    
     This value of the potential
comes from the jet parameters deduced for 3C 303 
with ${\cal Z}=90\Omega$ (Kronberg et al. 2011).
   (Note that magnetic insulation breakdown 
can also occur {\it along} the
jet due to the reflected wave from a localized region of increased series 
inductance along the transmission line.)

     The voltage across the Poynting jet,
$V = \Delta V_+ +\Delta V_-$, is assumed
 to be made up of an incident component $\Delta V_+$
 propagating along the jet channel with velocity $u_\varphi$
 and a reflected component $\Delta V_-$ with velocity $-u_\varphi$.    
 The  corresponding jet current is    $I= \Delta I_+ +\Delta I_-$.
    The spatial profile of the incident wave 
$\Delta V_+$  is considered to be a  smooth rise to a constant value (taken to be unity) as shown in the top panel of Fig. 5.
       Such jet onsets are observed in the form of the sporadic
formation of  parsec-scale components of radio galaxies and quasars
(e.g., Zensus et al. 1998).
   The onset may be explained by the global magnetic instability
of the disc (Lovelace et al. 1994).

    In such cases the wave amplitudes can be written as
\begin{equation}
\Delta V_\pm(z,t)=\int_{-\infty}^\infty d\omega~ \Delta V_\pm(\omega)
\exp[i\omega(\pm z/u_\varphi -t)]~,
\end{equation}
with $\Delta V_\pm(\omega)^*=\Delta V_\pm(-\omega)$ and
with
\begin{equation}
\Delta V_- (\omega)={{\cal Z}_L -{\cal Z} 
\over {\cal Z}_L +{\cal Z} }~\Delta V_+(\omega)
\end{equation}
where ${\cal Z}_L(\omega)$ is in general frequency-dependent
with real and imaginary parts.
   Figure 5 shows a case where ${\cal Z}_L=R-i\omega {\cal L} $ is
due to a resistance $R$  assumed equal to ${\cal Z}$ in series with and inductance ${\cal L}$.   An estimate of ${\cal L}$ is simply
$r_2 L$, where $r_2$ is the jet radius and $L$ is given by
equation (9).

\begin{figure}
\centering
\includegraphics[scale=.4]{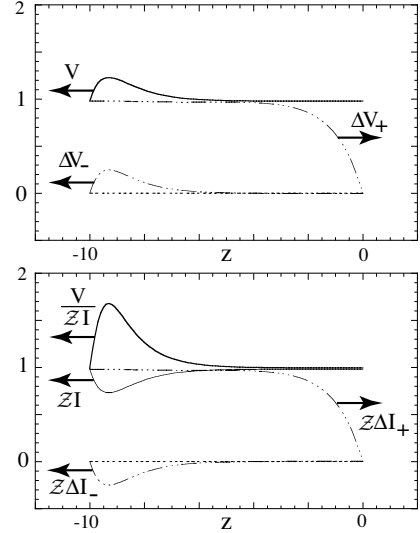}
\caption{Reflection of an incident wave 
($\Delta V_+,~\Delta I_+$)
off of a stationary load of impedance ${\cal Z}_L$
and the reflected wave ($\Delta V_-,~\Delta I_-$) for the 
circuit in Fig. 4.   
 The net voltage is
 $V = \Delta V_+ +\Delta V_-$ and the net current  
 $I= \Delta I_+ +\Delta I_-$.  The top panel shows 
 the voltages and the bottom panel the currents as
 well as the  ratio $V({\cal Z}I)^{-1}$.  
   If this ratio is larger than a critical value, ${\cal C}$ (equation 20),
then the magnetic insulations will breakdown.
 For this example, ${\cal Z}_L={\cal Z} -i\omega {\cal L}$, where
 ${\cal L}$ is the inductance of the load. }
\end{figure}

    The top panel of Fig. 5 shows that there is a {\it backward}
propagating positive   voltage wave  $\Delta V_-$.
The rise-time of the wave is equal to that of the incident wave while
the decay-time is equal to the $L$ over $R$ time of the load.
   The voltage wave is positive because the inductance of the
load appears initially to the incident wave as a
high impedance.
At the same time the reflected current $\Delta I_-$ is
negative.   The net voltage and current are such
that there is a {\it backward} propagating wave with
$V/({\cal Z}I) >{\cal C}$.   This corresponds
to a breakdown of the magnetic insulation.

     The condition $V/({\cal Z}I) >{\cal C}$ can
exist only transiently (for a time $ < r_2/c$) 
because for  $E_r <0$  the insulation breakdown 
will lead to a radial inflow of positive
charge with speed $c(1-{ B}^2/{E}^2)^{1/2}$  
and an opposite radial outflow of negative charge.  At
the same time both positive and negative charges
drift in the ${\bf E \times B}$ direction with speed
$c|B/E|$.
    This will bring the transmission line to
the threshold condition where
$V/({\cal Z}I) = {\cal C}$  or  $|{\bf E}| =
|{\bf B}|$.

    Both positive and negative charged particles in the  region
where $|{\bf E}| =
|{\bf B}|$  are accelerated
 in the ${\bf E} \times {\bf B}$ direction which
is approximately the $+z$ direction.  
     The acceleration is highly efficient in that
a particle exposed to the wave for a time $ \Delta t $
is accelerated to a Lorentz factor 
\begin{equation}
\gamma_{\rm E\times B} \approx 1 + {1\over 2}\left({6 e B \Delta t \over m c}\right)^{2/3}~,
\end{equation}
which is much larger than unity.   
 This formula is easily derived from the equations of motion.
     The electromagnetic  energy in the backward propagating wave 
is dissipated into kinetic energy of the accelerated particles.

    As a numerical example relevant to the HST-1 component
in M87, we consider $r_2= 5$  pc and $B = 0.5$ mG 
(Harris et al. 2009) and
assume that the jet consists of electrons and positrons.
    De Young (2006) discusses different models
of extra-galactic jets involving plasmas dominated
by  electrons and positrons,
electrons and ions, and electromagnetic fields (i.e., Poynting
flux jets). 
     For the electron-positron case,
\begin{equation}
\gamma_{\rm E\times B} \approx 1.2\times 10^7 \left({B \over 0.5{\rm mG}}\right)^{2/3}
\left( {\Delta t \over 5\times 10^8{\rm s}}\right)^{2/3}~,
\end{equation} 
where the time $\Delta t$ is normalized to $r_2/c \approx 5\times 10^8$s. 
    The accelerated particles will in a time $\sim r_2/c$ run into the
region of the load  where the magnetic field has a spread of directions.
   The $L/R$ time  $\sim 5\times 10^{10}$s is much longer
than the dissipation time for the backward propagating wave. 
     The synchrotron radiation peak from these high-$\gamma$ leptons
is at $\nu_{\rm syn} = 3e B \gamma^2/2m_ec \approx 3 \times 10^{17}{\rm Hz}$ or a photon energy $\approx 1300$ eV.       

       If the electromagnetic
radiation of the source $\dot{\cal E}_{\rm rad}$ is entirely
due to the accelerated particles, then
$\dot{\cal E}_{\rm rad} = 
\dot{N}_\ell m_ec^2 \gamma_{\rm E\times B}$, 
where ${N}_\ell$ is the number of accelerated leptons.  
    Estimating
$\dot{N}_\ell \sim \pi r_2^2c n_\ell$
and taking $\dot{\cal E}_{\rm rad}=10^{42}$erg/s (Harris et al. 2009) gives the lepton number density 
$n_\ell \sim 4.5 \times 10^{-9}~{\rm cm}^{-3}$.
    From this we obtain the  so called `plasma beta',
$\beta_p = 4\pi n_\ell \gamma_{\rm E\times B} m_e c^2/B^2
\sim 2.2$.   This implies that the load region of the jet is
{\it not force-free} and that it
will expand.  
Prior to the acceleration of
the particles when the power flow in the jet is
unidirectional, $\beta_p=1.9\times 10^{-7}\Gamma$  so
that the jet is dominantly {\it force-free}.
     Synchrotron and other losses will give rise to a power law
distribution of lepton energies $\propto \gamma^{-q}$ with
$q\geq 2$ and with a cutoff at  $\gamma_{\rm E\times B}m_e c^2$. 

     The estimated Lorentz factor (24) is 
larger than a critical value denoted $\gamma_{\rm c}$ set by the 
condition that the inverse Compton scattering of a lepton off
 a synchrotron photon (energy $\varepsilon_{\rm syn}$), 
which gives an inverse Compton photon 
(energy $\varepsilon_{\rm IC}\approx \gamma^2\varepsilon_{\rm syn}$),
be such that the subsequent scattering
of the synchroton and secondary inverse Compton photons is above the
threshold for electron-positron pair production 
[$\varepsilon_{\rm syn}\varepsilon_{IC} \approx (m_e c^2)^2$].   
     This condition gives 
\begin{equation}
\gamma_{ c} =\left({2m_e c^2 \over 3\hbar \omega_c}\right)^{1/3} \approx
3.9\times 10^5~,
\end{equation}
where $\omega_c = eB/m_e c$ is the non-relativistic cyclotron frequency
(e.g., Lovelace 1987).
At this Lorentz factor, $\varepsilon_{\rm syn} =1.3$ eV and $\varepsilon_{\rm IC}= 2\times 10^{11}$eV.

     The acceleration of particles is more complicated in the case
of an electron-ion jet.   The accelerated electrons
will tend to increase $-J_z$ more than the opposite effect from the
ions.  This will give rise to an inductive ambipolar 
$E_z$ field.  It in turn retards the electron acceleration and enhances the ion acceleration.
However,  analysis of this problem is beyond the scope
of the present work.

\section{Conclusions}

    We have  developed a model of Poynting jets as a  transmission line carrying a net axial current and  having   a potential drop  across it.
        The currents and voltages are of the order of
 $ 3\times 10^{18}$ A and $2.7\times 10^{20}\beta$ V.
    The energy of ions accelerated across this voltage is 
large enough  to account for ultra high energy cosmic
rays (Lovelace 1976;  Biermann et al. 2001; Ostrowski 2002).
   Further, we derive the transmission line impedance and
the electromagnetic energy flow in the jet.
   The observed current in 3C 303 is used to independently
estimate the electromagnetic
energy flow in this magnetically dominated jet.
    Time-dependent but not necessarily small
perturbations of a Poynting-flux jet -  possibly triggered
by an irregularity in the jet -
are described by the  ``telegraphers' equations,'' which 
are wave equations for the current and voltage on the line. 
   The voltage and current consist in general of forward and backward propagating components.
   The disturbance of a  Poynting jet by an irregularity  can
give rise to localized dissipation in the jet which may explain the 
enhanced synchrotron radiation in the knots of the
3C 303 jet and in the much smaller apparently stationary knot HST-1.

  Lastly, we  consider  relativistic Poynting jets relevant to
parsec-scale jets such as HST-1.   The reflected voltage
wave from an inductive load (or jet termination) can lead to a backward
propagating wave which causes the magnetic insulation to breakdown.
That is, it gives $|{\bf E}| / |{\bf B}|\geq 1$.   
   At the threshold for breakdown, $|{\bf E}| / |{\bf B}|=1$,
positive and negative particles are directly accelerated 
in the ${\bf E \times B}$ direction which is approximately
along the jet axis.  
     Particles can be accelerated up  
to  Lorentz factors $\sim 10^7$ in a short time interval of the
order of the light travel time across the jet.
   This particle acceleration mechanism is distinct from
particle acceleration in  shock waves and that in magnetic field
reconnection.

   The breakdown of magnetic insulation $|{\bf E}| / |{\bf B}|\geq 1$
is not possible in a plasma that is  modeled everywhere by ideal relativistic
magnetohydrodynamics (RMHD).  This is because the Ohm's
law of RMHD ${\bf E+ v\times B}/c =0$ requires that 
$|{\bf E}|/|{\bf B}|  \leq |{\bf v}|/c<1$ everywhere.   Of course an
actual plasma    can readily  have $|{\bf E}| /|{\bf B}|\geq 1$.  
A common example is magnetic reconnection where there is a region in which the direction of the magnetic field reverses so that $|{\bf B}|$ goes through zero on  a surface.  
       But even in regions of non-zero $|{\bf B}|$ a
plasma can have $|{\bf E}| / |{\bf B}|\geq 1$ transiently as in pulse
propagation along magnetically insulated transmission lines (e.g., Shope et al. 1978; Samokhin 2010).
     The restrictive Ohm's law constraint of RMHD can be avoided in
relativistic-electromagnetic particle-in-cell simulations where the
orbits of individual particles are calculated 
(e.g., Lovelace, Gandhi, \& Romanova 2005).f

      An open question regarding the propagation of
relativistic current-carrying Poynting jets is the kink instability. 
      In the context of  laboratory current-carrying plasmas,  the kink instability is predicted and observed to occur if the
 current is larger than a critical value which is the Kruskal-Shafranov
 condition (e.g., Kadomtsev 1966; Huarte-Espinosa et al. 2012).   The theory of the
 instability for relativistic Poynting jets has not been developed.
     Observational evidence for large current flows in astrophysical jets
(e.g., Kronberg et al. 2011) suggest that the jets can withstand
the kink instability.  On the other hand a mechanism for the nonlinear stabilization of the kink instability was proposed by Kadomtsev (1966, p. 188).

\section* {Acknowledgments:}  
 
 We thank  D.E. Harris, G.S. Bisnovatyi-Kogan,  S. Dyda, and M.M. Romanova for valuable discussions.     Also, we thank the referee
 for valuable criticism.
 RVEL was supported in part by NASA grants  NNX10AF-63G and NNX11AF33G and by NSF grant AST-1008636. PPK acknowledges support from NSERC Canada Discovery Grant A5713.

\appendix
\section{Waveguide Model of Poynting Jet}

     There are a wide range of axisymmetric,
translationally-symmetric, time-independent Poynting jet models
which satisfy the relativistic, force-free equation,
\begin{equation}
 {d B_z^2 \over dr}+ {1\over r^2}{d [ r^2(B_\phi^2 -E_r^2)] \over dr} =0~,
 \end{equation}
(cgs units).   This assumes that there is plasma everywhere, but that
the `plasma beta'  - the ratio of the kinetic energy-density to
electromagnetic field energy-density is $\ll 1$.   
    The time-independence of the solutions of (A1) implies that
the energy flow in the jet  is {\it unidirectional}.  Here, we
assume this is the $+z$ direction.
    Realistic models can in principle be derived for example
from relativistic particle-in-cell simulations (e.g., Lovelace et al. 2005).
    Here, we discuss a waveguide-like jet model  alternative to that considered in
the text.   The electric and magnetic fields are similar to the fields in a
TEM coaxial waveguide.
This model is suggested by 
the work of Bisnovatyi-Kogan \& Lovelace (1995).

     Equation (A1) is satisfied by the fields shown in
Fig. A1.    For $r < r_1$, the 
magnetic field is $B_z=$ const and the electric field is zero.  
For $r_1 <r <r_2$, the magnetic
field is $B_\phi =-B_0(r_1/r)<0$, and the electric field is 
$E_r=\eta B_\phi<0$ where $B_0= (1-\eta^2)^{-1/2}B_z$, 
and $\eta=$ const $<1$ is dimensionless (cgs units).
 For $r_2<r_2<r_3$,
the magnetic field is $B_\phi=(1-\eta^2)^{1/2}B_0(r_1/r)$ and
the electric field is zero.   The condition on the magnetic field at $r_3$
is the same as discussed in the text.   The plasma in the
region $r_1<r<r_2$ has a uniform axial ${\bf E \times B}$
drift velocity $\eta c \hat{\bf z}$ or Lorentz factor $\Gamma=
(1-\eta^2)^{-1/2}$.
    This model involves only a small number of parameters,
$(r_1, ~r_2, ~B_0,~ \eta)$.

\begin{figure}
\centering
\includegraphics[scale=.4]{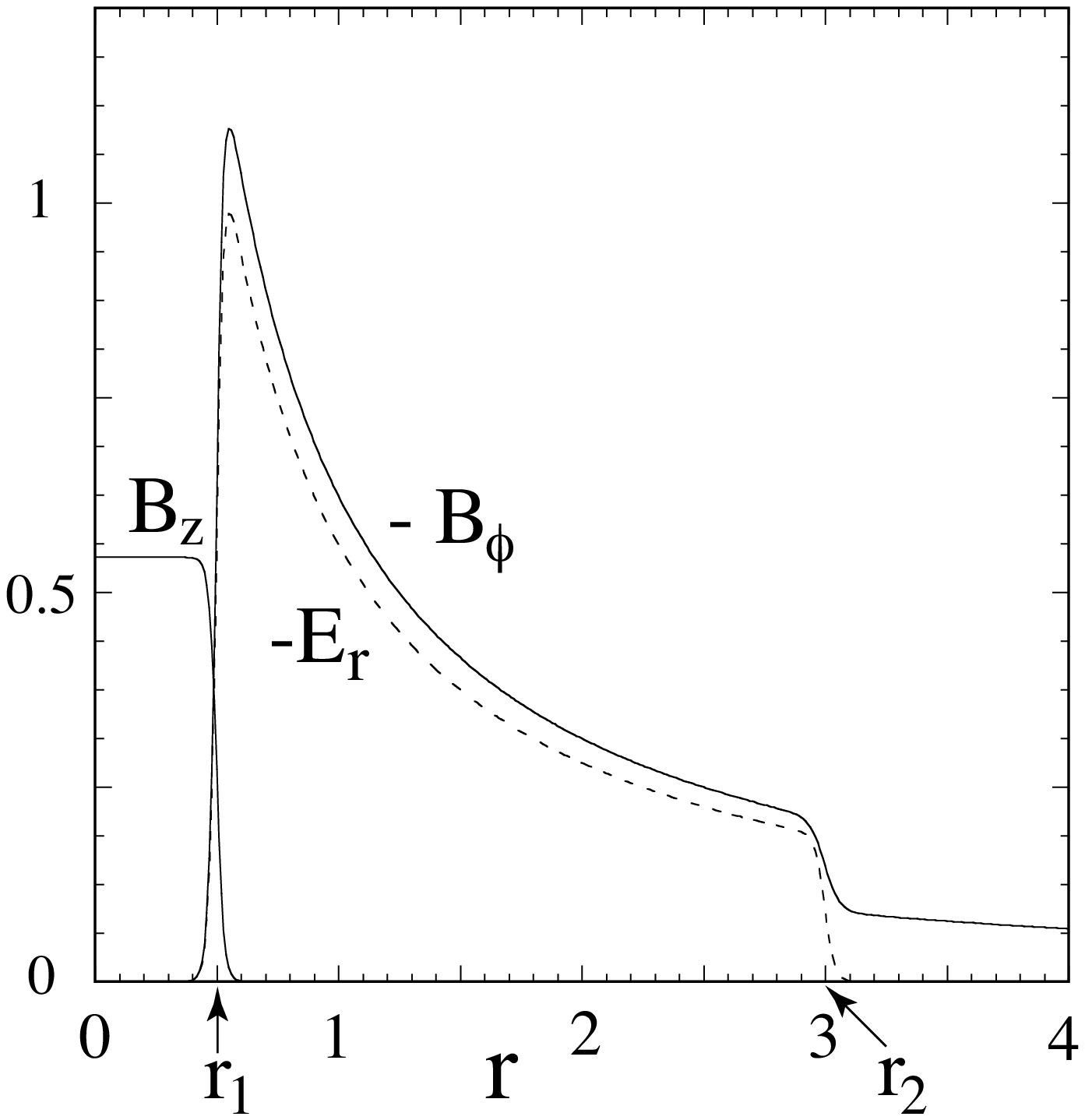}
\caption{Field profiles of a Poynting jet model alternative to 
that of the text (eqn. 2 and Fig.  2).    Here, 
we have taken $\eta=0.9$,  $r_1=0.5$ with a narrow boundary layer, and
$r_2= 3$ also with a narrow boundary layer.}
\end{figure}

      In MKS units,  the current carried by the core of the jet (in A) is
 $I_0 = 2\pi r_1 B_0/\mu_0$.
       Thus the  poloidal flux carried by the jet is
\begin{equation}
\Phi_z = \pi r_1^2 B_z = {1\over 2}(1-\eta^2)^{1/2} \mu_0 r_1 I_0~.
\end{equation}
The voltage drop across the jet is
\begin{equation}
V_0 =-\int_{r_1}^{r_2} dr E_r = {\cal Z}_0 I~,
\end{equation}
where
\begin{equation}
{\cal Z}_0={1\over 2\pi}\sqrt{\mu_0\over \epsilon_0} \eta \ln\left({r_2\over r_1}\right) =60~ \eta \ln\left({r_2\over r_1}\right)~\Omega~[{\rm MKS}]~.
 \end{equation}
 This formula is the counterpart of equation (4). 
    The electromagnetic power flow in the Poynting jet is
\begin{equation}
\dot{E}_j = {2 \pi \over \mu_0}\int_{r_1}^{r_2} r dr~E_rB_\phi
= {\cal Z}_0 I_0^2~,
\end{equation}    
which agrees with the equation (3).

 The electric field energy per unit length is
\begin{equation}
w_E ={\epsilon_0 \over 2}2\pi \int_{r_1}^{r_2} r dr E_r^2~ = ~{1\over 2}C V_0^2~,
\end{equation}
where
\begin{equation}
C={2\pi\epsilon_0 \over  \ln(r_2/r_1)}~
\end{equation}
is the capacitance per unit length of the jet.

    The magnetic energy per unit length of the jet is
\begin{eqnarray}
w_B&=&{\pi\over  \mu_0}\int_0^{r_1} r dr  B_z^2+
{\pi \over \mu_0} \int_{r_1}^{r_3} r dr B_{\phi}^2~,
\nonumber \\
&=&{1\over2}L I_0^2~.
\end{eqnarray}
     Carrying out the integrals we find
\begin{equation}
L ={\mu_0 \over 2 \pi}\left[{1-\eta^2\over 2}+\ln\left({r_2\over r_1}\right)+
(1-\eta^2)\ln\left({r_3\over r_2}\right)\right],
\end{equation}   
which is the inductance per unit length of the jet.

    From equations (A7) and (A9) we find the wave speed
on the jet is
\begin{equation}
u_\varphi =c\left(1+{1-\eta^2 \over 2\ln(r_2/r_1)}  +{(1-\eta^2) \ln(r_3/r_2)
\over \ln(r_2/ r_1)}\right)^{-1/2}~,
\end{equation}
The wave impedance of the transmission line is
\begin{equation}
{\cal Z} ={\cal Z}_0\left(1+{1-\eta^2 \over 2\ln(r_2/r_1)}  +{(1-\eta^2) \ln(r_3/r_2)\over \ln(r_2/ r_1)}\right)^{1/2}~.
\end{equation}
Equations (A10) and (A11) are the analogues of equations
(12) and (14).

\end{document}